\documentclass[aps,prl,showpacs,twocolumn,floats,epsfig,pdflatex]{revtex4}
\usepackage{amssymb}
\usepackage{amsbsy}
\usepackage{amsmath}
\usepackage{epsfig}
\usepackage{graphicx}
\newcommand\beq{\begin{equation}}
\newcommand\eeq{\end{equation}}
\newcommand\bea{\begin{eqnarray}}
\newcommand\eea{\end{eqnarray}}
\newcommand\al{\alpha}

\newcommand\de{\delta}
\newcommand\ep{\epsilon}
\newcommand\si{\sigma}
\newcommand\dg{\dagger}
\newcommand\pa{\partial}

\newcommand\non{\nonumber}
\newcommand\bib{\bibitem}

\begin{document}

\title {Spin injection into a metal from a topological insulator}

\author{S. Modak$^{(1)}$, K. Sengupta$^{(1)}$, and Diptiman Sen$^{(2)}$}

\affiliation{$^{(1)}$Theoretical Physics Department, Indian Association for
the Cultivation of Science, Jadavpur, Kolkata 700 032, India \\
$^{(2)}$Center for High Energy Physics, Indian Institute of Science,
Bangalore 560 012, India}
\date{\today}

\begin{abstract}
We study a junction of a topological insulator with a thin
two-dimensional (2D) non-magnetic or partially polarized
ferromagnetic metallic film deposited on a 3D insulator. We show
that such a junction leads to a finite spin current injection into
the film whose magnitude can be controlled by tuning a voltage $V$
applied across the junction. For ferromagnetic films, the direction
of the component of the spin current along the film magnetization
can also be tuned by tuning the barrier potential $V_0$ at the
junction. We point out the role of the chiral spin-momentum locking
of the Dirac electrons behind this phenomenon and suggest
experiments to test our theory.
\end{abstract}


\maketitle

Topological insulators (TI), a class of three-dimensional (3D)
insulators with strong spin-orbit coupling, are known to possess
gapless Dirac-like quasiparticles on their surfaces whose existence
originates from the special topological properties of their bulk
bands \cite{zhang1,hassan1,kane1,kane2,teo,qi1,exp2,expt1,hassan2}. Such
insulators, which are essentially 3D generalizations of their 2D
counterparts which exhibit the quantum spin-Hall effect, have attracted
a lot of theoretical and experimental attention in recent years.
These TIs can be classified as strong or weak depending on their
sensitivity to time reversal symmetric perturbations. The surfaces
of the strong TIs have an odd number of Dirac cones; the number
and positions of these cones depend on the nature of the surface
concerned \cite{kane2,zhang1,hassan1}. The odd number of the Dirac
cones ensures that any surface impurity which conserves time
reversal symmetry does not destroy the low-energy Dirac properties
of the quasiparticles on the surface. For several compounds such as
$\rm Bi_2 Te_3$ and ${\rm Bi_2 Se_3}$, specific surfaces have been found
with a single Dirac cone near the $\Gamma$ point of the 2D surface
Brillouin zone \cite{hassan1,exp2,hassan2}.

A Dirac cone on the surface of a TI is described by the Hamiltonian
\bea H_{\hat n} [v_F] &=& \int \frac{dk_i dk_j}{(2\pi)^2}
~\psi^\dg_{\vec k} ~(\hbar v_F {\hat n} \cdot {\vec \si} \times
{\vec k} - \mu I) ~\psi_{\vec k}, \label{ham1} \ \eea
where $\vec \si (I)$ denotes the Pauli (identity) matrices in spin space,
${\hat n}$ denotes the unit vector normal to the TI surface which hosts
electrons with momentum components $k_i$ and $k_j$, $v_F$ is the Fermi
velocity, $\mu$ is the chemical potential, and $\psi= (\psi_{\uparrow},
\psi_{\downarrow})^T$ is the annihilation operator for the Dirac
spinor \cite{kane4}. In this notation, $\uparrow (\downarrow)$
denotes components of the quasiparticle spin along (opposite to)
${\hat z}$. Recently, several novel features of these surface Dirac
electrons have been studied. These include the existence of Majorana
fermions in the presence of a magnet-superconductor interface on the
surface \cite{kane4,been1}, generation of a state resembling a $p_x
+ i p_y$-wave superconductor but with time reversal symmetry via proximity
to a $s$-wave superconductor \cite{kane4}, anomalous magnetoresistance of
ferromagnet-ferromagnet junctions \cite{tanaka1}, realization of a magnetic
switch in junctions of these materials \cite{mondal1}, spin textures with
chiral properties \cite{hassan2}, control of spin transport and polarization
using gate voltages and electric fields \cite{burkov,garate}, realization
of a Lifshitz transition in thin TI films \cite{zyuzin}, and spin-polarized
STM spectra \cite{das1}. However, junctions of such TIs with conventional
metals and ferromagnets have not been studied so far.

\begin{figure} \includegraphics[width=\linewidth]{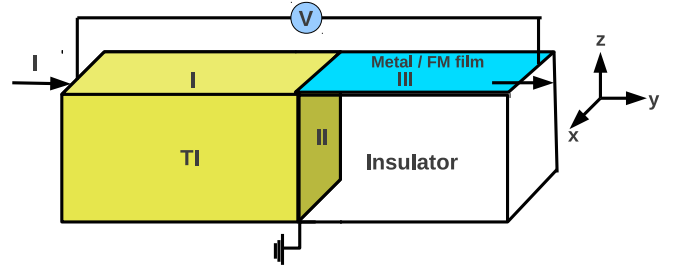}
\caption{(Color online) Schematic picture of the proposed junction.
The two interfaces I and II of the topological insulator have
low-energy quasiparticles with Dirac-like properties while the
non-magnetic/ferromagnetic metallic thin film III deposited atop the
insulator has conventional electrons obeying the Schr\"odinger
equation. See text for details.} \label{fig1} \end{figure}

In this Letter, we study the transport properties of a junction of a
TI with a conventional insulator with a non-magnetic or partially
polarized ferromagnetic metallic film deposited on it as
schematically shown in Fig.\ \ref{fig1}. The regions I and II in
Fig.\ \ref{fig1} refer to the two surfaces of the TI which host
chiral Dirac quasiparticles, while region III consists of
conventional electrons obeying the Schr\"odinger equation. We show
that due to the chiral spin-momentum locking of the Dirac electrons
on the surfaces of the TI, the transport through such a junction can
lead to a finite spin current in the film without any skew
scattering which is necessary in conventional spin-Hall materials to
generate such currents. The magnitude of the spin current generated
can be controlled by a voltage $V$ applied across the junction {\it
without the involvement of any external magnetic field}. We also
study the dependence of this spin current on the barrier potential
$V_0$ at the junction and show that it displays an oscillatory
feature with a decaying envelope as a function of $V_0$. This
behavior draws from both the chiral nature of the Dirac
quasiparticles in regions I and II which leads to the oscillatory
nature of the spin current \cite{comment1}, and the presence of the
conventional Schr\"odinger electrons in region III which leads to an
exponential decay of the spin current with increasing $V_0$.
Finally, we demonstrate that for ferromagnetic films, the direction
of the component of the spin current along the film magnetization
($J_z$) can be controlled by tuning $V_0$. To the best of our
knowledge, the generation of spin current using the chirality of the
Dirac quasiparticles on the surface of the TIs {\it whose direction
and magnitude can both be controlled electrically} has not been
proposed before; we therefore expect our proposal to generate
significant interest in the field of spintronics.

We begin with the analysis of the junction in Fig.\ \ref{fig1} when
$V_0=0$. In region I, the Hamiltonian for the Dirac quasiparticles
is given by Eq.~(\ref{ham1}) with ${\hat n}={\hat z}$. The wave
function for these quasiparticles with a transverse momenta $k_x$
and energy $\ep=eV$ moving along $\pm {\hat y}$ can be obtained by solving
the Dirac equation $H_{\hat z} [v_1] \psi_1 = \ep \psi_1$ and is
given by $\psi_{\pm} = [1, -i \exp(\pm i \theta_k)]^T \exp[i(k_x x
\pm k_y y)]/\sqrt{2}$, where $\theta_k = \arccos[\hbar v_1 k_x/(\ep
+\mu)]$, $\ep= -\mu + \hbar v_1 \sqrt{k_x^2+k_y^2}$, and $v_1$ is
the Fermi velocity of the quasiparticles in region I. The wave
function of the Dirac quasiparticles in region I can thus be written as
\bea \psi_{I} &=& \psi_+ + r \psi_-, \label{wav2} \eea
where $r$ is the reflection amplitude. We note that for any incident angle
$\theta$, $\psi_+$ and $\psi_-$ have $\langle \psi_+ |\si_x |
\psi_+ \rangle = \sin(\theta_k) = - \langle \psi_- |\si_x |
\psi_-\rangle$. Thus the reflected Dirac quasiparticle has the
opposite spin orientation along $\hat x$ compared to its incident
counterpart. This phenomenon is reminiscent of Andreev reflection
from a superconducting junction where the charge and the transverse
momenta of the reflected quasiparticle change sign. In contrast, for
the junctions considered here, the transverse in-plane component of
the quasiparticle spin changes sign upon reflection. In what
follows, we shall show that this spin reversal is at the heart of
the generation of a finite spin current in region III.

In region II, ${\hat n}={\hat y}$, and the wave function of the
Dirac quasiparticles moving along $-{\hat z}$ with transverse
momenta $k_x$ and energy $\ep$ can be obtained by solving $H_{\hat
y} [v_2] \psi_{2} = \ep \psi_2$ and is given by
\bea \psi_{II}= t_1 \psi_2, \quad \psi_2 = [u_{k},v_{k}]^T ~ e^{i(-k_z z +k_x
x)}/\sqrt{2}, \label{wav3} \eea
where $u_{k}[v_k]= \sqrt{1+[-]\cos(\phi_k)}$, $\phi_k = \arccos[\hbar v_2
k_x/(\ep + \mu)]$, $\ep= -\mu + \hbar v_2 \sqrt{k_z^2+k_x^2}$, $v_2 = \beta^2
v_1$ is the Fermi velocity, and $t_1$ denotes the transmission
probability of the Dirac quasiparticles in region II. In the rest of
this work, we shall choose $\beta =\sqrt{v_2/v_1} \le 1$.

In the metallic film (region III), the Hamiltonian for the electrons
can be written as $H_{III}= \hbar^2(k_x^2+k_y^{'2})/(2m) - \mu - A
\si_z$, where $\mu$ and $m$ are the chemical potential and mass of
the electrons in the film, and $A$ is proportional to the magnetization
of the electrons. For a non-magnetic film $A=0$, while for a fully
polarized ferromagnetic film $A \to \infty$. In what follows, we
first consider a non-magnetic film for which $A=0$. The wave
function of the electrons in region III is then given by
\bea \psi_{III} &=& [t_2,t_3]^T ~e^{i(k_x x+k'_y y)} /\sqrt{2},
\label{wav4} \eea where $\ep= -\mu + \hbar^2(k_x^2+k_y^{'2})/(2m)$,
and $t_2$ and $t_3$ denote the transmission amplitudes of spin-up
and spin-down electrons in region III.

The boundary condition on these wave functions involves continuity
of current through the junction which yields
\bea v_1 \psi_{I}^\dg \si_x \psi_{I} - v_2 \psi_{II}^\dg \si_x \psi_{II} &=&
\frac{\hbar}{m} Im (\psi_{III}^\dg \pa_y \psi_{III}), \label{bc1} \eea
where it is understood that all fields are evaluated at the
junction line $y=0$ (for $\psi_I$ and $\psi_{III}$) and $z=0$ (for
$\psi_{II})$. We note that the unusual boundary condition (Eq.\
\ref{bc1}) in which a current without derivatives in regions I and
II (Dirac equation) has to be matched with a current involving a
first derivative in region III (Schr\"odinger equation) is generic for any
junction involving a TI and a conventional material. (The situation here is
different from a junction of ordinary materials with spin-orbit coupling
where the current involves a first derivative on both sides \cite{molenkamp}).
Below, we present a general solution to this problem with (discussed in
Eq.~(\ref{bc4}) below) and without a barrier potential \cite{comment2}.

In the absence of any barriers at the junction, the general solution of
Eq.~(\ref{bc1}) is given by two linear conditions on the wave functions
\cite{suppl},
\beq \psi_{III} = c (\psi_I + \beta \psi_{II}), ~~~~\frac{\hbar}{m} \pa_y
\psi_{III} =\frac{iv_1 \si_x}{c} [\psi_{I} -\beta \psi_{II}], \label{bc2} \eeq
where $c$ is an arbitrary real constant; we will set $c=1$ for simplicity.
Note that the metal/ferromagnet decouples from the TIs for $c \to 0$ or
$\infty$. Substituting Eqs.~(\ref{wav2}-\ref{wav4}) in Eq.\ \ref{bc2}, we
obtain the following relations between $r$, $t_1$, $t_2$, and $t_3$,
\bea 1+r+(-)\beta u_k t_1 &=& t_2 (\al t_3), \non \\
e^{i \theta_k} + r e^{- i \theta_k} +(-) i \beta v_k t_1 &=& it_3 (i
\al t_2), \label{bc3} \eea
where $\al= \hbar k'_y/(mv_1)$. Solving for $r$, $t_{1,2,3}$ from
Eq.~(\ref{bc3}), we get
\bea r &=& {\mathcal N}/{\mathcal D}, \quad t_1 = 2(1-\al^2)\sin(\theta_k)/
(\beta {\mathcal D}), \label{sol1} \\
t_2 &=& \frac{4}{\mathcal D} \sin(\theta_k)(u_k+\al v_k), \quad t_3=
\frac{4}{\mathcal D} \sin(\theta_k)(v_k+\al u_k), \non \eea
where ${\mathcal N} = -iu_k [\exp(i \theta_k) (1+\al^2) - 2i \al] - v_k [
(1+\al^2)+2i \al \exp (i \theta_k)]$, and ${\mathcal D} = i
[u_k(1+\al^2)+ 2\al v_k] \exp(-i\theta_k) + v_k(1+\al^2)+2\al u_k$.

\begin{figure} \includegraphics[width=\linewidth]{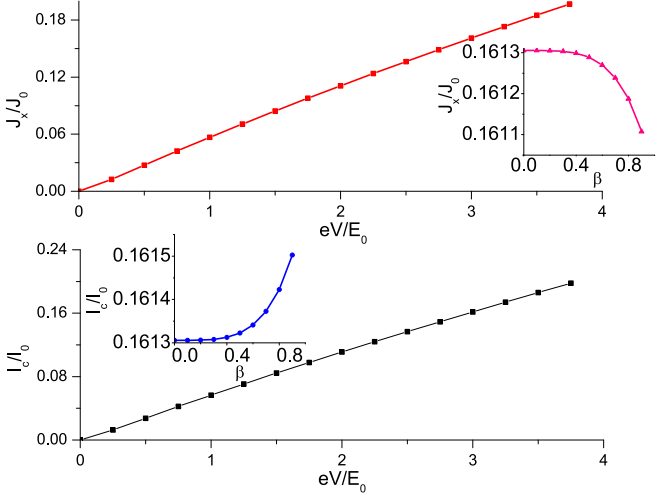}
\caption{(Color online) Plot of $J_x$ (top panel) and $I_c$ (bottom panel) in
a non-magnetic metallic film as a function of $V$ for $\beta=1$. The insets
show their $\beta$ dependence for $eV/E_0=3$.} \label{fig2} \end{figure}

Eq.~(\ref{sol1}) predicts a net spin current along $\hat x$ in the
metallic region: $J_x = (\hbar v_1/2) \sum_{k_x} \langle
\psi_{III}|\al \si_x|\psi_{III}\rangle = (\hbar v_1/2) \sum_{k_x}
\al (t_2^{\ast} t_3 +{\rm h.c.})$. This can be written as
\beq J_x = \frac{8 J_0}{\pi} \int_{-E/(2E_0)}^{E/(2E_0)} dx
~\frac{\al[(1+\al^2)\sin(\phi_k) + \al] \sin^2(\theta_k)}{|{\mathcal
D}|^2}, \label{spinc1} \eeq
where $J_0 = \hbar v_1 k_0$,
$x=k_x/k_0$, $k_0= m v_1/\hbar$, $E=eV+\mu$, and $E_0= \hbar v_1
k_0/2$. A plot of $J_x/J_0$ as a function of the applied voltage
$eV/E_0$, shown in the bottom panel of Fig.\ \ref{fig2}, confirms
that the net spin current is finite and its amplitude depends on $V$.
The inset shows the dependence of the spin current on $\beta$ for a
fixed $V$ and confirms the presence of a finite $J_x$ for the entire
range of $0 < \beta \le 1$. The charge current is given by $I_c=ev_1
\sum_{k_x} \al (|t_2|^2+|t_3|^2)$ which can be written as
\beq I_c = \frac{4I_0}{\pi} \int_{-E/(2E_0)}^{E/(2E_0)} dx ~\frac{\al[1+\al^2
+ 2 \al \sin(\phi_k) ] \sin^2(\theta_k)}{|{\mathcal D}|^2}, \eeq
where $I_0 = e v_1 k_0$. The top panel of Fig. 2 shows the
variation of $I_c/I_0$ with the applied voltage $V$, while the inset
depicts its variation with $\beta$. We note that the charge current
displays a qualitatively similar behavior as the spin current along
$\hat x$. The net spin current along $y$ and $z$ vanishes. For $J_z$
this can be seen by noting that $u_k \to v_k $ under $k_x \to -k_x$.
Consequently $t_2 \to t_3$ under this transformation which leads to
a zero net value for $J_z \sim \sum_{k_x} \alpha (|t_2|^2 -|t_3|^2)$.
Also, using the fact that $t_2$ and $t_3$ are both purely imaginary
(Eq.~(\ref{sol1})), it can be easily shown that $J_y=0$ \cite{comment3}.

\begin{figure} \includegraphics[width=\linewidth]{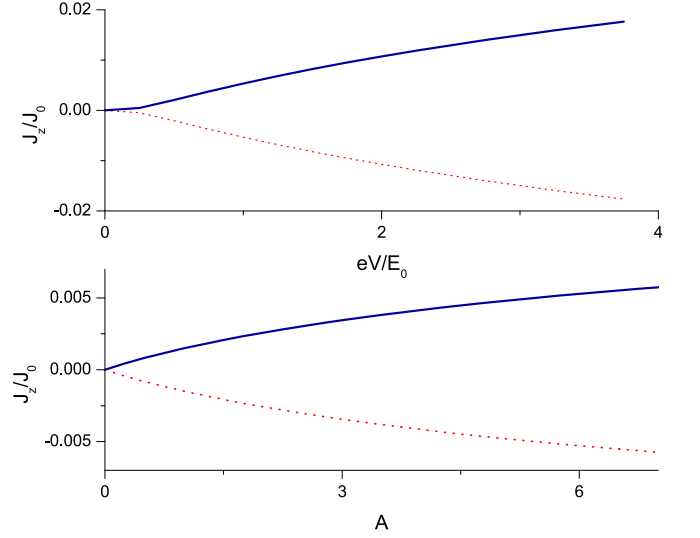}
\caption{(Color online) Plot of $J_z$ versus $eV$ when $\mu=A$ (red
dashed line) and $\mu=-A$ (blue solid line) for $A/E_0=3$ (top
panel) and as a function of $A$ (bottom panel) for $\beta=1$ and
$eV/E_0=3$ and $\mu=A$ (red dashed line) and $\mu=-A$ (blue solid line).
} \label{fig3} \end{figure}

Next, we address the behavior of the spin and charge currents for a
ferromagnetic film where $A\ne 0$. In this case, the wave function
in region III is given by \bea \psi_{III}^{\rm FM} &=& [ t_2 e^{i
k^{(1)}_y y}, t_3 e^{i k^{(2)}_y y}]^T~ e^{i k_x x} /\sqrt{2},
\label{wavfm}\eea where $k^{(1)[(2)]}_y = \left[ 2m(\ep+\mu +[-]
A)/\hbar^2 -k_x^2\right]^{1/2}$. The boundary condition on the wave
function is given by Eq.~(\ref{bc2}) and yields \bea t_{2[3]} &=&
-4i e^{i \theta_k} \sin(\theta_k) (u_k [\alpha_2 u_k] + \al_1 v_k
[v_k])/{\mathcal D_1}, \label{solmag} \eea where $\al_{1(2)} = \hbar
k^{(1)}_y(k^{(2)}_y)/(mv_1)$, and ${\mathcal D_1}= p_1 - ip_2 \exp(i
\theta_k)$, where $p_1 = u_k(1+\al_1 \al_2) + 2 v_k \al_2$, and
$p_2= v_k (1+\al_1 \al_2) + 2 u_k \al_2$. Note that for $A=0$,
$\al_1=\al_2 = \al$; in this limit Eq.~(\ref{solmag}) matches with
Eq.~(\ref{sol1}). Using this wave function, it is straightforward to
compute the expression for $I_c$ following the method outlined
earlier which shows qualitatively similar behavior as in metallic
films.

The key difference between the ferromagnetic and the non-magnetic films which
we now focus on is that the former films allow a non-zero $J_z$. This is
easily seen from Eq.\ (\ref{solmag}) by noting that $t_2(k_x) \ne t_3(-k_x)$
due to the difference of velocities of the up and the down spin
quasiparticles; this leads to a finite $J_z$ in region III given by
\bea J_z ~=~ \frac{J_0}{2\pi}~ \int_{-E/(2E_0)}^{E/(2E_0)} dx ~(\alpha_1
|t_2|^2-\alpha_2|t_3|^2). \label{jzcurrent} \eea
In what follows, we set $\mu =A$ or $\mu=-A$ so that the spin-up or spin-down
Fermi surface is aligned with the Fermi surface in region I. Increasing
$A$ therefore pushes the other Fermi surface away from the Fermi
surface of region I and hence increases the effective spin
polarization of the film in region III. The behavior of $J_z$,
computed by substituting Eq.\ (\ref{solmag}) in Eq.\ (\ref{jzcurrent}),
is shown in Fig.\ \ref{fig3}. The top panel of the figure shows the
variation of $J_z$ as a function of $eV/E_0$ for $A/E_0=3$, while the
bottom panel shows the dependence of $J_z$ on $A$ for fixed $eV/E_0=3$
and $\beta=1$. These plots demonstrate the presence of a finite $J_z$ for a
large range of $V$; the sign of $J_z$ depends on the magnetization of the
film while its magnitude can be controlled by the applied voltage $V$.

Next, we consider the effect of a finite barrier potential $V_0$ applied over
a region $d$ at the junction. In what follows we will consider the
thin-barrier limit where $V_0 \to \infty$ and $d\to 0$ keeping $\chi= V_0
d/(\hbar v_1)$ finite. In this limit, the boundary condition to be imposed
on the wave functions reads \cite{suppl,dsen}
\bea \psi_{III} &=& e^{-i\chi \si_x}\psi_I + \beta e^{i\chi \si_x/\beta^2}
\psi_{II}, \label{bc4} \\
\frac{\hbar}{m v_1} \pa_y \psi_{III} - 2 \chi  \psi_{III} &=& i
\si_x (e^{-i\chi \si_x} \psi_I - \beta e^{i\chi \si_x/\beta^2}
\psi_{II}). \non \eea Substituting Eqs.~(\ref{wav2}), (\ref{wav3}),
and (\ref{wavfm}) in Eq.~(\ref{bc4}), we again obtain a set of four
equations for $r$ and $t_{1,2,3}$ which reads \bea \hspace*{-.4cm}
t_{2[3]} &=& \frac{1}{\sqrt{2}}(a_1 [ia_2] + r a_1^{\ast}
[ia_2^{\ast}] +[-] \beta t_1 b_1 [ b_2]), \non \\
\hspace*{-.4cm} (\al_{1[2]} + 2 i \chi) t_{3[2]} &=& \frac{1}{\sqrt{2}}(a_1
[-ia_2] + r a_1^{\ast} [-ia_2^{\ast}] - \beta t_1 b_1 [ b_2]) \nonumber\ \eea
where $a_1 = \cos(\chi)-\sin(\chi)\exp(i \theta_k)$, $a_2 = \sin(\chi) +
\cos(\chi) \exp(i \theta_k)$, and $b_{1[2]} = u_k[v_k] \cos(\chi/\beta^2) +
i v_k[u_k] \sin(\chi/\beta^2)$. The solution of these equations yields
\bea t_{2[3]} &=& [-2(b_{1[2]}+b_{2[1]}(\al_{1[2]}+2i\chi))
(a_1^*a_2-a_2^*a_1)]/{\mathcal D}_2, \nonumber \\ \label{sol2} \eea
where ${\mathcal D_2} = [-2 b_1(2\chi + i\alpha_1) + b_2\{-1+(\alpha_1-2i
\chi)(\alpha_2-2i \chi)\}]i a_1 + [b_1(1+(\al_1+2i\chi)(\alpha_2
+2i\chi))+2b_2(\al_2+2i\chi)]a_2^*$ for magnetic films. The
corresponding expressions for the non-magnetic films can be obtained
by putting $\alpha_1=\alpha_2$ in Eq.\ (\ref{sol2}). We note that Eq.\
(\ref{sol2}) reproduces Eq. (\ref{solmag}) for $\chi=0$.

{}From Eq.~(\ref{sol2}), we find that the barrier potential $\chi$
enters the transmission amplitudes $t_{2[3]}$ both through the
$\cos(\chi)$ and $\sin(\chi)$ factors in $a_{1(2)}$ and $b_{1(2)}$
leading to an oscillatory $\chi$ dependence of $t_{2(3)}$ and
through the appearance of $\chi$ in the numerator and denominator of
Eq.~(\ref{sol2}), which, in the limit of large $\chi$, leads to a
decay of $t_{2(3)}$ with increasing $\chi$. The former behavior
arises from the Dirac nature of the electrons in regions $I$ and
$II$, while the latter is a consequence of the conventional
Schr\"odinger nature of the electrons in region III. Consequently,
we expect $t_{2(3)}$ to have an oscillatory dependence on $\chi$ along
with a decaying envelope. We note that such a behavior is different
from what is found in analogous junctions involving solely
Dirac or solely conventional materials.

\begin{figure} \includegraphics[width=\linewidth]{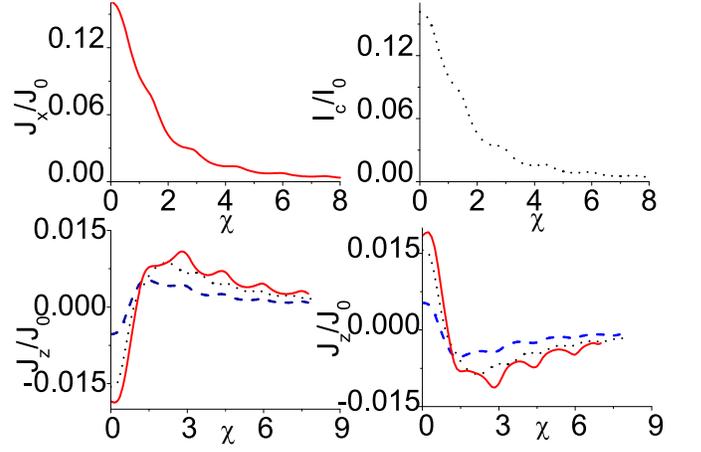}
\caption{(Color online) Top left (right) panel: Variation of $J_x$
($I_c$) with $\chi$ in a metallic film for $eV/E_0 = 3$. Bottom left
(right) panel: Variation of $J_z$ with $\chi$ in a ferromagnetic
film for $\mu=A=3E_0$ ($\mu=-A=3E_0$) for $eV/E_0=4$ (red solid
line), $3$ (black dotted line) and $1$ (blue dashed line). For all
the plots $\beta=1$. 
} \label{fig4} \end{figure}

To compute the spin and charge currents, we substitute the values of
$t_{2(3)}$ from Eq.~(\ref{sol2}) in the expressions $J_z = \hbar v_1
\sum_{k_x} (\alpha_1|t_2|^2-\alpha_2|t_3|^2)$ for ferromagnetic
films, and $I_c = ev_1 \sum_{k_x} \alpha_1 ( |t_2|^2+|t_3|^2)$ and
$J_x = \hbar v_1 \sum_{k_x} \alpha (t_3^{\ast} t_2 + {\rm h.c.})$
for non-magnetic films. The resulting dependence of $J_z$, $I_c$ and
$J_x$ on $\chi$ is shown in Fig.\ \ref{fig4} for $\beta=1$. The top
panel shows the behavior of $J_x$ and $I_c$ for non-magnetic films
for $eV/E_0=3$. We find that both $J_x$ and $I_c$ display small
oscillatory features with an overall monotonic decay as a function
of $\chi$. We have found qualitatively similar behavior of $J_x$ and
$I_c$ in magnetic films. In contrast, $J_z$ for magnetic films,
shown in the left (right) bottom panels of Fig.\ \ref{fig4} for
$\mu=A$ ($\mu=-A$) for several representative values of $eV/E_0$
displays a non-monotonic behavior with increasing $\chi$. In
particular, these plots demonstrate that the sign of the spin
currents gets reversed with increasing $\chi$ which allows for the
possibility of tuning the sign of $J_z$ by tuning $V_0$. This leads
to electrical control of both the magnitude and the direction of
$J_z$ in magnetic films via tuning externally applied voltages $V$
and $V_0$.

To experimentally verify our theory, we propose measuring the
current in region III for magnetic films using a ferromagnetic
contact. When the direction of magnetization of the contact is along
${\hat z}$, it will measure only the current due to the spin-up
electrons: $I_{\uparrow} = ev_1 \sum_{k_x} \al_1 |t_2|^2= [I_c + e
J_z/(2\hbar)]/2$. Similarly a contact with magnetization along
$-{\hat z}$ will record $I_{\downarrow} = [I_c - eJ_z/(2\hbar)]/2$.
Thus the difference between these two currents will provide a
measure of the spin current through the film as a function of the
applied voltage $V$ and the potential barrier $V_0$: $J_z = 2\hbar
(I_{\uparrow}-I_{\downarrow})/e$. Thus such an experiment can verify
the predicted dependence of $J_z$ on $V$ and $V_0$ \cite{suppl}.

In conclusion, we have studied a junction of a TI with a thin
metallic/partially polarized ferromagnetic film deposited over an
ordinary insulator. For ferromagnetic films, we have shown that such
a junction can be used to generate a finite spin current along $\hat
z$ whose magnitude and direction can both be controlled by externally
applied voltages without the presence of any external magnetic field.
Our work shows that the chiral spin-momentum locking of the Dirac
quasiparticles on the surfaces of the TI is at the heart of this
phenomenon. Finally, we have suggested a simple experiment to test
our theory.

K.S. and D.S. thank DST, India for financial support through grants
SR/S2/CMP-0001/2009 and SR/S2/JCB-44/2010 respectively.

\section{Appendix A}
\label{appa}

In this Section, we provide some supplemental material to the main
text related to the derivation of the boundary condition Eq.\
\ref{bc4} and numerical estimate of the measured current in
suggested experiment.

\subsection{Current conserving boundary conditions at a junction}

In this section, we will find the general time reversal invariant
boundary condition which satisfies the current conservation relation
at the junction discussed in our paper. We begin with the
Hamiltonians in the three regions. Region I of the topological
insulator (TI) is defined by the half-plane $z=0$ and $y<0$, and has
the Dirac Hamiltonian \beq H_I ~=~ i \hbar v_1 ~[- \si_x \pa_y ~+~
\si_y \pa_x]. \eeq Region II of the TI, given by the half-plane
$y=0$ and $z<0$, has the Hamiltonian \beq H_{II} ~=~ i \hbar v_2 ~[-
\si_z \pa_x ~+~ \si_x \pa_z]. \eeq Region III of the non-magnetic
metal/ferromagnet thin film is defined by the half-plane $z=0$ and
$y>0$, and has the Schr\"odinger Hamiltonian \beq H_{III} ~=~ -
\frac{\hbar^2}{2m} ~(\pa_x^2 ~+~ \pa_y^2) ~-~ \mu ~-~ A \si_z. \eeq
The time evolution equations $i\hbar \pa \psi_a /\pa t = H_a \psi_a$
($a=I,II, III$) are invariant under time reversal ($t \to -t$ and
complex conjugation of all numbers) if $\psi_i \to \si_y \psi_i^*$
and $A=0$.

The junction of the three regions is given by the line $y=z=0$. The
conservation of the total current coming into the junction from the
three regions is given by Eq. (5) of our paper, namely, \bea && v_1
(\psi_{I}^\dg \si_x \psi_{I})_{y=0-} ~-~ v_2 (\psi_{II}^\dg \si_x
\psi_{II})_{z=0-} \non \\
&=& \frac{\hbar}{m} Im (\psi_{III}^\dg \pa_y \psi_{III})_{y=0+}.
\label{cond} \eea In order to satisfy this equation, let us assume
linear relations between the wave functions at the junction of the
form \bea (\psi_{III})_{y=0+} &=& A_1 (\psi_I)_{y=0-} ~+~ A_2
(\psi_{II})_{z=0-},
\non \\
\frac{\hbar}{m} ~(\pa_y \psi_{III})_{y=0+} &=& i \si_x ~[A_3
(\psi_I)_{y=0-} ~
+~ A_4 (\psi_{II})_{z=0-}], \non \\
&& \label{rel1} \eea where the $A_i$ are four parameters. The
relations in Eq.~(\ref{rel1}) will be time reversal invariant if the
$A_i$ are real. We can now check that Eq.~(\ref{cond}) will be
satisfied if $A_1 A_3 = v_1$, $A_2 A_4 = - v_2$ and $A_1 A_4 + A_2
A_3 = 0$. This implies that the $A_i$ can be written in terms of a
single real parameter $c$ as $A_1 =c$, $A_2 = c \beta$, $A_3 =
v_1/c$ and $A_4 =-v_1 \beta/c$, where $\beta = \sqrt{v_2/v_1}$; this
gives \bea (\psi_{III})_{y=0+} &=& c [(\psi_I)_{y=0-} + \beta
(\psi_{II})_{z=0-}], \non \\ \frac{\hbar}{m} (\pa_y
\psi_{III})_{y=0+} &=& \frac{iv_1 \si_x}{c} ~[
(\psi_{I})_{y=0-} - \beta (\psi_{II})_{z=0-}], \non \\
&& \label{rel2} \eea which is Eq. (6) of our paper. In the limits
$c\to 0$ (or $\infty$), we obtain $\psi_{III} = 0$ (or $\pa_y
\psi_{III} = 0$); in either case, the current into the junction from
region III vanishes, so that the metal/ferromagnet gets decoupled
from the two TI regions. The value of $c$ in a given system will
depend on its microscopic details such as an underlying lattice
model. For the numerical calculations in our paper, we have simply
set $c=1$.

The above analysis assumed that there is no barrier present at the
junction. A realistic system may be expected to have some potential
barriers present at the junction in all three regions. Let us assume
thin barriers of the form $V (y) = V_0$ for $-d < y < 0$ in region
I, $V (z) = V_0$ for $-d < z < 0$ in region II, and $V (y) = V_0$
for $0 < y < d$ in region III. For simplicity, we have assumed the
barrier width ($d$) and height ($V_0$) to be the same in all three
regions; we will eventually be interested in the $\de$-function
limit $d \to 0$ and $V_0 \to \infty$ keeping $dV_0/(\hbar v_1) =
\chi$ constant. In Ref. \onlinecite{sen}, it has been shown that
$\de$-function barrier in a Dirac Hamiltonian produces a
discontinuity in the wave function of the form
\bea (\psi_I)_{y=0-} &=& e^{-i\chi \si_x} ~(\psi_I)_{y=-d}, \non \\
(\psi_{II})_{z=0-} &=& e^{i\chi (v_1/v_2) \si_x}
~(\psi_{II})_{z=-d}. \label{rel3} \eea A $\de$-function barrier in a
Schr\"odinger Hamiltonian produces no discontinuity in the wave
function (i.e., $(\psi_{III})_{y=d} = ( \psi_{III})_{y=0+}$), but
there is a discontinuity in the first derivative of the form \beq
\frac{\hbar}{m} ~[(\pa_y \psi_{III})_{y=d} ~-~ (\pa_y
\psi_{III})_{y=0+}]~ =~ 2 \chi v_1 (\psi_{III})_{y=d}. \label{rel4}
\eeq Substituting Eqs.~(\ref{rel3}-\ref{rel4}) in Eq.~(\ref{rel2}),
and setting $c=1$, we obtain \bea & & (\psi_{III})_{y=d} ~=~
e^{-i\chi \si_x} (\psi_I)_{y=-d} + \beta
e^{i(\chi/\beta^2) \si_x} (\psi_{II})_{z=-d}, \non \\
& & \frac{\hbar}{m} (\pa_y \psi_{III})_{y=d} - 2 \chi v_1
(\psi_{III})_{y=d}
\non \\
& & =~ iv_1 \si_x [e^{-i\chi \si_x} (\psi_I)_{y=-d} -\beta
e^{i(\chi/\beta^2) \si_x} (\psi_{II})_{z=-d}]. \eea In the limit $d
\to 0$, this gives Eq. (14) of our paper.

\subsection{Numerical estimates of measured currents}

In this section we provide estimates for $I_{\uparrow}$ and
$I_{\downarrow}$ which are to be measured in the proposed
experiment. For a typical TI surface, the group velocity of the
Dirac electrons turns out be $ v_1 \sim 10^6$m/s. This give us an
estimate of $k_0 = mv_1/\hbar \simeq 8.63 \times 10^{-9} {\rm
m}^{-1}$. Using this, one finds $I_0 = ev_1 k_0 \simeq 13.83$mA and
$E_0= mv_1^2/2 \simeq 2.89$eV. Since $I_{\uparrow}$ and
$I_{\downarrow}$ can be written as \bea I_{\uparrow (\downarrow)}
&=& \frac{4I_0}{\pi} \int_{-E/(2E_0)}^{E/(2E_0)} ~dx ~\alpha_1
|t_2|^2 ~(\alpha_2 |t_3|^2), \eea we find numerical values of
$I_{\uparrow} = 0.113$mA and $I_{\downarrow}= 0.0906$mA for $\chi=0$
and $\mu=-A = 3 E_0$. This indicates that the visibility $V$ defined
by \bea V &=& \left|\frac{I_{\uparrow} -I_{\downarrow}}{I_{\uparrow}
+ I_{\downarrow}}\right| \eea is close to $0.1$ which means that
such current measurements are well within the reach of current
experimental standards. For finite and large barrier strength
$\chi=5$, the corresponding numbers are $I_{\uparrow}=0.0068$mA and
$I_{\downarrow}=0.012$mA which leads to $V \simeq 0.28$.

\end{document}